\documentclass[fleqn]{annalen}
\usepackage{graphics}
\begin{document}
\newcommand{\volume}{11}              
\newcommand{\xyear}{2000}            
\newcommand{\issue}{5}               
\newcommand{\recdate}{15 November 1999}  
\newcommand{\revdate}{dd.mm.yyyy}    
\newcommand{\revnum}{0}              
\newcommand{\accdate}{dd.mm.yyyy}    
\newcommand{\coeditor}{ue}           
\newcommand{\firstpage}{507}         
\newcommand{\lastpage}{510}          
\setcounter{page}{\firstpage}        
\newcommand{\keywords}{cosmology, high energy physics} 
\newcommand{\PACS}{95.35+d, 98.80.Cq}
\newcommand{\shorttitle}
{P. Brax et al., Cosmological implications of the Supergravity Tracking 
Potential} 
\title{Cosmological Implications of the Supergravity Tracking Potential}
\author{P.\ Brax$^{1}$ and J.\ Martin$^{2}$} 
\newcommand{\address}
  {$^{1}$Service de Physique Th\'eorique, CEA-Saclay, 
  F-911191 Gif/Yvette Cedex, 
  \\ \hspace*{0.5mm} France\\ 
  $^{2}$DARC, Observatoire de Meudon UMR 8629, 
  92195 Meudon Cedex, 
   \\ \hspace*{0.5mm} France}
\newcommand{\email}{\tt martin@edelweiss.obspm.fr} 
\maketitle
\begin{abstract}
  It is demonstrated that any realistic model of quintessence should 
be based on Supergravity since, when the scalar field is on tracks 
today, $Q\approx m_{\rm Pl}$. This improves the 
agreement between theoretical predictions and the current 
observations. In particular, a generic property is that ratio 
$\omega _{\rm Q}\equiv p_{\rm Q}/\rho _{\rm Q}$ is pushed towards $-1$. A 
string-inspired model is proposed where the potential is given by $V(Q)=
\frac{\Lambda ^{4+\alpha }}{Q^{\alpha }}e^{\frac{\kappa }{2}Q^2}$. The 
model predicts $\omega _{\rm Q}\approx -0.82$, a value less that one sigma 
from the current likehood value.
\end{abstract}

\section{Introduction}
\label{Intro}
There are now several hints indicating that our Universe could 
presently undergo a phase of accelerated expansion. These hints 
consist in a set of recent observations including particulary 
(but not only) measurements of the Hubble law using type Ia 
supernovae \cite{SNIa}. All these observations seem to point towards the
same 
conclusion: a fluid with a negative pressure to energy density ratio 
contributes for $70\%$ of the matter content of the Universe and therefore 
represents the dominant component of this matter content.
\par
If this conclusion is confirmed, this immediately raises the question of 
the physical nature and origin of this fluid. At first sight, a 
natural candidate is the cosmological constant. A cosmological constant 
is equivalent to a fluid with an equation of state given by $p=-\rho$.
This 
seems in agreement with the recent analysis of the data which 
indicate that $-1<p/\rho <-0.8$ or $-1<p/\rho <-0.6$ according to 
Refs.~\cite{E} and \cite{WCOS}. However, $\Omega _{\Lambda }\approx 0.7$
corresponds 
to an energy scale of $\approx 5.7 \times 10^{-47} \mbox{GeV}^4$ which is 
very far from the natural scales of High Energy Physics. Therefore, even 
if there is presently no contradiction with observations, it seems that 
this hypothesis runs into theoretical problems.
\par
Another explanation was recently put forward in Refs.~\cite{SWZ}. It
consists 
in assuming that the unknown fluid is a scalar field named quintessence.
Then, 
the following questions need to be answered. Firstly, 
why is it so that this fluid is dominating today? This is the coincidence
problem. 
Secondly, in order to explain $\Omega _{\rm Q}\approx 0.7$, is it 
necessary to fine tune a free parameter of the theory to a value very far
from 
the natural scales of Particle Physics? In that case, nothing would have
been gained 
in comparison with the cosmological constant case. This is the 
fine tuning problem. Thirdly, is it possible to find values of the 
pressure to energy density ratio compatible with recent findings? This 
is the equation of state problem. Fourthly, can realistic quintessence
models be 
implemented in the realm of High Energy Physics? This is the 
model building problem.
\par
It has been argued in Refs.~\cite{SWZ} that these four problems can be
partially 
tackled if the potential is given by:
\begin{equation}
\label{potsusy}
V(Q)=\frac{\Lambda ^{4+\alpha }}{Q^{\alpha }},
\end{equation}
where $\Lambda $ and $\alpha >0$ are free parameters. The coincidence
problem 
is solved because the equation of motion of the scalar field possesses a 
tracking solution. Just after reheating, at a 
redshift of $z\approx 10^{28}$, the allowed initial conditions for the
energy 
density of the quintessence field are $10^{-37} \mbox{GeV}^4<\rho _{\rm Q}
<10^{61} \mbox{GeV}^4$ (i.e. $100$ orders of magnitude!). For any initial
value 
in this range, the field is led to the same solution. The fine tuning
problem 
is partially solved. For $\alpha =11$, for example, $\Lambda \approx 
10 ^{10}\mbox{GeV}$ in order to have $\Omega _{\rm Q}\approx 0.7$.
Therefore 
the natural scale of the problem is now comparable to the natural scales 
of Particle Physics. However, it is still necessary to adjust this free 
parameter. The main problem for the models based on the 
potential (\ref{potsusy}) could come from the equation of state. It has 
been show that $\omega _{\rm Q}\equiv p_{\rm Q}/\rho _{\rm Q}$ cannot be 
less than $-0.7$. This seems in disagreement with observations if one 
believes in the estimates of Ref.~\cite{E}. However, according to
Ref.~\cite{WCOS}, 
there is still an open windows for these models. Finally, the question of
the 
model building have been adressed, in particular, in Ref.~\cite{PB} where
a 
model based on global supersymetry (SUSY) has been proposed. In this
model, the 
K\"ahler potential is flat, $K(Q,Q^*)=QQ^*$, and the superpotential is
given 
by $W(Q)=\Lambda ^{3+a}/Q^a$. This leads to an inverse power law scalar 
potential.
\par
The aim of this article is to show that any realistic model of
quintessence 
must be based on Supergravity (SUGRA). We proove that, generically, taking 
into account SUGRA improves the agreements with observations, especially 
with regards to the equation of state. In order to 
illustrate these general properties, we exhibit a specific model where 
concrete calculations can be performed and we propose a new potential 
for the quintessence field, the supergravity tracking potential. 

\section{Taking into account SUGRA}

The questions evoked in the previous section can be adressed if the 
field is on tracks. According to Ref.~\cite{SWZ}, this means that 
it should satisfy the equation:
\begin{equation}
\label{attractor}
\frac{{\rm d}^2V(Q)}{{\rm d}Q^2}=\frac{9}{2}
\frac{\alpha +1}{\alpha }(1-\omega _{\rm Q}^2)H^2,
\end{equation}
where $H$ is the Hubble constant. This equation implies 
$Q\approx m_{\rm Pl}$ now. Since SUGRA corrections are of the order 
$Q/m_{\rm Pl}$, they are crucial for quintessence and any realistic model 
should be based on SUGRA.
\par
The SUGRA ($N=1$, $D=4$) scalar potential is given by $V=\kappa
^{-2}e^G(G^iG_i-3)+V_{\rm D}$, 
where $\kappa \equiv 8\pi/m_{\rm Pl}^2$, $G\equiv \kappa K+\ln (\kappa
^3\vert W\vert ^2)$ 
and $V_{\rm D}\ge 0$ is a term coming from the gauge sector. Then, we can
draw 
general conclusions. Firstly, the K\"ahler and super potentials should be
chosen 
such that negative contributions do not dominate the scalar potential.
This is 
not the case if the SUGRA corrections to the model proposed in
Ref.~\cite{PB} are 
taken into account. Secondly, we see that for a typical polynomial
K\"ahler potential $K$, the term 
$e^G$ is unimportant throughout almost all the cosmic evolution since 
$Q\ll m_{\rm Pl}$. This means that the term $G^iG_i$ should be responsible 
for the tracking properties. Thirdly, the term $e^G$ should dominate now 
since we have $Q\approx m_{\rm Pl}$. Since the exponential is a rapidly 
growing function, this implies that the potential energy should largely 
dominate the kinetic energy today. In other words, this automatically 
pushes the ratio $\omega _{\rm Q}$ towards $-1$ which is precisely 
needed in order not to be in conflict with observations. These properties 
are generic, i.e.~they do not depend on the details of the model, even if
of 
course it is certainly possible to find very specific cases where they are 
not true\footnote{In particular a logarithmic K\"ahler potential for
string moduli has been ruled out \cite{BM2}}.
\par
In Refs.~\cite{BM1,BM2},  string-inspired K\"ahler and superpotentials
were 
proposed. They lead to the SUGRA tracking potential given by:
\begin{equation}
\label{potsugra} 
V(Q)=\frac{\Lambda ^{4+\alpha }}{Q^{\alpha }}e^{\frac{\kappa }{2}Q^2},
\quad \alpha >11.
\end{equation}
In the next section, we investigate in more details the cosmological 
implications of this potential.

\section{Cosmological implications}

\begin{figure}
\centerline{\resizebox{7.25cm}{4.5cm}{\includegraphics{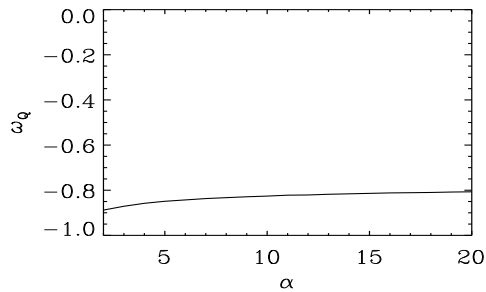}}}
\caption{$\omega _{\rm Q}-\alpha $ relation for the SUGRA tracking 
potential}
  \label{omega-alpha}
\end{figure}

In order to study the properties of the SUGRA tracking solutions, we 
have numerically integrated the full Einstein equations. 
\begin{figure}
\centerline{\resizebox{7.25cm}{4.25cm}{\includegraphics{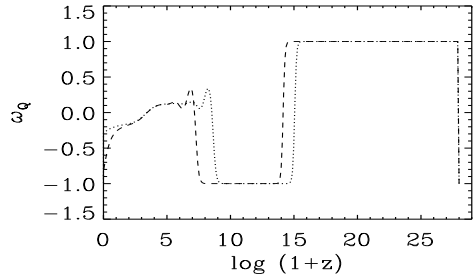}}}
\caption{The dotted line represents the evolution of $\omega _{\rm Q}$ 
for the potential given in Eq.~(\ref{potsusy}) with $\alpha =11$ whereas 
the dashed line represents the evolution of $\omega _{\rm Q}$ for the 
SUGRA tracking potential, Eq.~(\ref{potsugra}), with the same value 
of $\alpha $.}
  \label{omegaz}
\end{figure}
We have checked that the coincidence problem is still solved in the SUGRA
case, 
see Refs.~\cite{BM1,BM2}. The evolution of $\omega _{\rm Q}$ is displayed 
in Fig. (\ref{omegaz}). The influence of 
the exponential term at the end of the evolution, for small redshifts, is 
clearly visible. For $\alpha =11 $, the value of the equation of state 
for the potential given by Eq. (\ref{potsusy}) is $\omega _{\rm Q}\approx 
-0.29$, a value ruled out by observations. For the SUGRA tracking 
potential, we find:
\begin{equation}
\label{omega}
\omega _{\rm Q} \approx -0.82,
\end{equation}
for $\Omega _{\rm Q}\approx 0.7$. According to Refs. \cite{E}, this is 
less than one sigma from the likehood value. For $\Omega _{\rm Q}$ 
between $0$ and $1$, $\omega _{\rm Q}$ changes between $-0.22$ and 
$-0.995$. If $\Omega _{\rm Q}\approx 0.75$ then $\omega _{\rm Q}\approx 
-0.86$. Furthermore, these numbers do not require any fine tuning 
of the free parameter $\alpha $. This is because the value is mainly 
determined by the exponential factor which is $\alpha $ 
independent. This property is illustrated in Fig. (\ref{omega-alpha}).
\par
In conclusion, we would like to emphasize that taking into account 
SUGRA is mandatory if one wishes to construct a realistic 
model of quintessence. This is because when the field is on tracks, 
$Q\approx m_{\rm Pl}$. Then, a natural consequence is that 
theoretical predictions fit better the current data. In particular, 
the model presented here predicts $\omega _{\rm Q}\approx -0.82$ 
which lies into the one sigma error interval.

\vspace*{0.25cm} \baselineskip=10pt{\small \noindent The authors would 
like to thank Martin Lemoine and Alain Riazuelo for useful discussions 
and help. }

\end{document}